\documentclass[11pt,twoside]{article}

\usepackage{asp2006}
\usepackage{natbib}

\def\gtsim{{_>\atop{^\sim}}}
\def\ltsim{{_<\atop{^\sim}}}

\def\ccm{cm$^{-3}$}
\def\rs{s$^{-1}$}
\def\pow#1#2{#1$\times$10$^{#2}$}
\def\msol{M$_{\odot}$}
\def\kms{km~s$^{-1}$}

\def\hi{H~{\sc I}}

\def\hh{H$_2$}

\def\hho{H$_2$O}

\def\ddo{D$_2$O}
\def\dds{D$_2$S}
\def\hhoe{H$_2^{18}$O}
\def\hhhop{H$_3$O$^+$}
\def\hcop{HCO$^+$}
\def\dcop{DCO$^+$}

\def\hhco{H$_2$CO}

\def\ddco{D$_2$CO}
\def\ddcs{D$_2$CS}
\def\meth{CH$_3$OH}
\def\chddoh{CHD$_2$OH}
\def\cdddoh{CD$_3$OH}

\def\cchh{C$_2$H$_2$}

\def\ammo{NH$_3$}
\def\nddd{ND$_3$}
\def\nddh{ND$_2$H}

\def\oo{O$_2$}

\def\hhhp{H$_3^+$}
\def\hhdp{H$_2$D$^+$}
\def\ddhp{D$_2$H$^+$}
\def\nnhp{N$_2$H$^+$}
\def\nn{N$_2$}

\def\aap{{A\&A}}

\def\apjl{{ApJ}}
\def\apjs{{ApJS}}

\begin{document}
\title{Recent Astrochemical Results on Star-Forming Regions}   
\author{Floris van der Tak}   
\affil{SRON Netherlands Institute for Space Research \\ Landleven 12, 9747~AD
  Groningen, The Netherlands; vdtak@sron.nl}    

\begin{abstract} 
  This review discusses recent results on the astrochemistry of (mostly
  high-mass) star-forming regions. After an introduction on the use of chemistry
  in astrophysics and some basic concepts of astrochemistry, specific results
  are presented.  Highlighted areas are the use of chemistry in the search for
  massive circumstellar disks, the interaction of molecular clouds with cosmic
  rays, and the feedback effects of protostellar irradiation on the parent
  molecular cloud.  The review concludes with a discussion of future
  observational opportunities.
\end{abstract}

\section{The use of chemistry in astrophysics}
\label{sec:use}

There are several ways in which knowledge of chemistry is helpful to gain a
better understanding of the Universe. The prime area of interest is the
formation of stars and planets, which occurs in cold dark clouds which require
observation at long (infrared and radio) wavelengths.
The line radiation of molecules is an essential part of this effort, because it
is the only probe of the kinematics of these clouds, and because it is the major
way to determine their temperatures, volume densities, and other conditions.
Successful use of molecular lines to derive physical parameters requires some
understanding of chemistry to predict which molecules may be abundant under
which conditions. This use may be called `passive' astrochemistry.

A more active way to use chemistry in astrophysics makes use of the dependence
of the molecular composition of the gas on parameters which are otherwise hard
to estimate. This use involves the construction of chemical models typically
containing thousands of reactions; \citet{wakelam:errors} have studied the
accuracy of such models.  First, the chemical composition of the gas in such
models usually depends on the time, so that observations of molecular lines may
be used to estimate the ages of star-forming regions
\citep{doty:hotcores}. Second, since the chemistry needs time to respond to
changing conditions, molecular abundances often contain some memory of the
source history. An example are the deuterium bearing molecules seen in hot
molecular cores, which must be remnants of a previous cold phase. Third, the
chemical composition of star-forming matter may give clues to the presence of
(energetic) radiation which is difficult or impossible to observe directly. It
is clear that chemistry is a significant help in the understanding of
astrophysical processes.

More than 130 molecules are known to exist in interstellar space, of which 36
are known outside our Galaxy and 10 are known in the solid state
\citep{gibb:ice}.
%
%
The eight new discoveries of 2007 (see www.cdms.de for the latest updates) may
be grouped in a few `threads':
(i) Complex organics (by which astronomers usually mean 4$^+$-atomic carbon
chains), which probe gas-grain interactions and serve as a link to pre-biotic
molecules; the most recent addition is C$_3$H$_6$ \citep{marcelino:propylene}.
(ii) Fluorine and phosphorus compounds (PO, HCP, CF$^+$), which help to determine
elemental abundances and the composition of dust grains in dense clouds
(\citealt{neufeld:cf+}; \citealt{agundez:hcp}; \citealt{tenenbaum:po}).
(iii) Negative ions, which are useful to measure the
electron fraction of molecular gas (\S~\ref{sec:zeta}).
(iv) Deuterated molecules, which are important as
tracers of very early (cold and dense) phases of star formation (\S~\ref{sec:filters}).

Due to space limitations, this review is not comprehensive, but biased toward
the interests of its author. 
For a recent general review of astrochemistry see the books edited by
\citet{lis:asilomar} and \citet{combes:paris}.
The physics and chemistry of regions of low-mass star formation are reviewed by
\citet{bergin:review} and \citet{ceccarelli:ppv}.
An overview of methods to derive molecular abundances from spectral line data
can be found in \cite{vdtak:alma}.
The formation of high-mass stars is reviewed by \citet{zinnecker:araa} and \citet{beuther:ppv},
and is of course the topic of the current volume.

\section{Basics of Astrochemistry}
\label{sec:basics}

The chemical composition of interstellar molecular clouds depends strongly on
the physical conditions, particularly the temperature and the radiation field.
This discussion is restricted to dense interstellar clouds, defined
theoretically as $n \gtsim 10^4$\,\ccm\ or observationally as $A_V \gtsim 3$, so
that photoprocesses can be ignored. In such clouds, four main types of
environments can be distinguished:

\paragraph{Cold gas} At $T \ltsim 100$\,K, the main type of reactions occurring
in the gas phase are ion-molecule reactions. An example is the proton transfer
from \hh\ to CO: CO + \hhhp\ $\to$ \hcop\ + \hh. 
This type of reaction usually does not have any activation barrier and usually proceeds
at about the `Langevin' rate of $\sim$10$^{-9}$\,\ccm\,\rs. The necessary ions are
produced in the interaction of \hh\ with cosmic rays (\S~\ref{sec:zeta}).
One major source of uncertainty are the rates and the branching ratios of
dissociative recombination reactions of ions with free electrons \citep{mitchell:recom}.

\paragraph{Warm gas} Reactions between two neutral species occur
if the gas is warm enough to overcome their activation barriers. The rates of
these reactions and the heights of their barriers can be difficult to measure or
predict, especially when radicals are involved.
A classic case is the reaction O + OH $\to$ \oo\ + H with a measured rate
constant of \pow{3.5}{-11}\,\ccm\,\rs\ at $T=40$\,K, well above the predicted
value \citep{xu:o2}.
This reaction determines the main oxygen reservoir in cold interstellar clouds,
which is difficult to measure because the fine structure lines of O$^0$ do not
probe cold gas and \oo\ only has weak quadrupole lines. 
The abundance of \oo\ recently measured with the Odin satellite towards the
$\rho$\,Oph cloud \citep{larsson:o2} is well below theoretical predictions.
Laboratory data at $T<40$\,K and observations with \textit{Herschel}
(\S~\ref{sec:future}) are needed for further progress.

\paragraph{Cold dust} In dense clouds, the surfaces of dust grains act as
catalysts for reactions that would not take place in the cold gas phase.
An important example is the formation of \hhco\ and \meth\ by
successive additions of H atoms to CO molecules. Recent laboratory experiments indicate
that this process is very efficient (\citealt{watanabe:meth}; \citealt{fuchs:meth}).
One uncertainty in modeling such processes is the roughness of the surface which determines
the mobility of the H and O atoms (e.g., \citealt{cuppen:h2}).
Depending on this parameter, grain surface chemistry may operate at temperatures
up to $\sim$100\,K \citep{cazaux:h2}, but this remains subject of discussion
\citep{herbst:grains}. 

\paragraph{Warm dust} When dust grains are heated by the radiation from young
stars or by interstellar shock waves, any ice layers will evaporate.
The evaporation temperature varies from $\approx$20\,K for volatile species such
as CO and \nn\ to $\approx$110\,K for the more refractive \hho\ molecule which
makes up the bulk of the ice mantle \citep{collings:desorption}.

Observations of dense molecular cores without embedded stars often show a
differentiation between CO and \nn: in the core centers, CO appears depleted
while \nn\ (traced by \nnhp\ and \ammo) remains in the gas phase (e.g.,
\citealt{tafalla:cores}). This behaviour cannot be due to the difference in
evaporation temperature between the two species which \citet{bisschop:ice} has
shown to be very small.
Alternatively, CO freeze-out removes the major destroyer of \nnhp, so that its
abundance rises toward the centers of pre-stellar cores \citep{aikawa:review},
but this effect does not quite explain the observations \citep{flower:depletion}.

Significant rearrangement of the ice layers may occur during the warm-up phase
of the ice before the actual evaporation, which may lead to the formation of
more complex molecules \citep{garrod:ice}. This rearrangement is a more likely
source of molecular complexity than gas-phase processes, the
preferred model of the 1990's.

\section{Chemical Filters}
\label{sec:filters}

The rate coefficients of many chemical reactions depend on the temperature.
If the dependence is very strong, a molecule may almost exclusively exist in
warm or cold gas.
In an astrophysical context, this behaviour may be used to trace regions of a
particular temperature, a concept known as a chemical filter. Three particular
cases are:

\paragraph{Cold gas: \boldmath \hhdp}

The \hhdp\ molecule is produced in the gas phase by the reaction of \hhhp\ with
HD. At $T\ltsim 20$\,K, the back reaction is very slow, and if in addition the
density is high ($\gtsim$10$^5$\,\ccm), the main destroyers of \hhdp, CO and O,
will freeze out onto dust grains. Under these circumstances, the \hhdp/\hhhp\
ratio may approach or even exceed unity, and further reaction to \ddhp\ and
D$_3^+$ may even occur \citep{roberts:h3+}.
High abundances of \hhdp\ measured in a few dense pre-stellar cores and of
\ddhp\ in one confirm these predictions
\citep{caselli:h2d+,belloche:cham,hogerheijde:b68,vastel:d2h+}. 
A survey of \hhdp\ in 12 dense molecular cores with and without embedded stars
clearly shows a decrease of the \hhdp\ abundance as the young star warms up its
surroundings \citep{caselli:survey}.
The \hhdp\ molecule thus acts as a filter for the cold dense gas at the centers
of pre-stellar cores where most other molecules are frozen onto dust, and is the
only probe of the kinematics in this phase (e.g., \citealt{vdtak:l1544}).

Regions of high-mass star formation tend to have lower degrees of deuterium
fractionation than their low-mass counterparts; see \citet{fontani:n2d+} for a
recent example. The implication is that the cold pre-stellar phase for regions
of massive star formation has a short duration compared with the low-mass case,
or that the ambient gas is warmer in high-mass than in low-mass regions. The
duration argument is supported by source counts \citep{garay:review}.

In recent years, several multiply deuterated molecules have been detected toward
dense molecular cores: \ddco, \ddcs, \nddh, \dds, \chddoh, \nddd, \cdddoh, and
\ddhp (see \citealt{ceccarelli:ppv} for references). Th latest addition to this
list, after extensive searches, is the discovery of interstellar \ddo\
\citep{butner:d2o}. The low fractionation of \hho\ compared with other molecules
suggests that deuterium enrichment is primarily a gas-phase process.
The likely origin of multiply deuterated molecules is transfer of deuterons from
\hhhp\ isotopologues at low temperatures ($\ltsim$20\,K), aided by transfer from
deuterated C\hhhp\ and \cchh$^+$ at higher temperatures \citep{roueff:d}.
The measured abundances of multiply deuterated molecules imply that the
freeze-out of molecules onto grains is slow, suggesting grain growth in
pre-stellar cores \citep{flower:grains}.

\paragraph{Warm gas: \boldmath \hho}

There are three formation routes for interstellar water.
At low temperatures, \hho\ is produced in the gas phase by dissociative
recombination of \hhhop, which itself derives from O by reactions with \hhhp\
and \hh. However, \hho\ is created much more efficiently on the surfaces of dust
grains by H atom addition to adsorbed O atoms.
The ice mantles may desorb from the grains if they are thermally heated to
$T\gtsim 100$\,K by nearby young stars, or through photodesorption in regions
with significant ultraviolet radiation \citep{hollenbach}.
At high temperatures ($\gtsim$250\,K), \hho\ is produced efficiently in the gas
phase through the reactions of O and OH with \hh, which have significant
barriers \citep{wagner:rates}.

Far from embedded young stars, dense molecular cloud thus have a background
level of \hho\ originating in \hhhop\ recombination and photodesorption of \hho\ ice;
it is this \hho\ which is picked up in large-scale maps of \hho\ emission \citep{melnick:swas}
although excitation effects may complicate the picture \citep{poelman:swas}.
Close to young stars, the \hho\ abundance rises steeply because of thermal ice
evaporation.  Even higher \hho\ abundances are reached in outflows, where gas is
shock-heated to several 100\,K and the neutral-neutral channel kicks in
\citep{franklin:outflows}. Because of these effects, \hho\ acts as a filter for
warm gas in star-forming regions.

One application of this filter is the search for massive circumstellar disks. 
High-mass stars may form through disk accretion like their low-mass
counterparts, perhaps with an increased accretion rate. The alternative model
where high-mass stars form through coagulation of lower-mass stars or
pre-stellar cores probably only applies to a minority of cases, as extremely
high stellar densities are required. However, positive evidence for accretion
disks around young high-mass stars has been hard to find, as reviewed by Q.\
Zhang (this volume). The main problem is confusion of the molecular line
emission from the disk with that from the surrounding envelope.

Observations of the \hhoe\ line at 203\,GHz with the Plateau de Bure
Interferometer have now revealed such a massive circumstellar disk
\citep{vdtak:h2o}. The disk radius is $\approx$400\,AU, the mass of
$\approx$0.8\,\msol\ is $\approx$5\% of the mass of the central star, and the
observed velocity gradient in the \hhoe\ line is consistent with the Keplerian
rotation speed.
Together with NGC 7538 IRS11 \citep{sandell:7538} and IRAS 20126
\citep{cesaroni:20126}, this source is one of the more compelling cases for an
accretion disk around a young high-mass star.

\paragraph{Shocked gas: SiO}

The star formation process entails gas parcels moving both inward and outward,
and shocks occur frequently. The shocked gas has its own chemistry, because the
gas is heated to $\sim$1000\,K, grain mantles are disrupted, and even grain
cores are shattered if the shocks are fast enough.  The erosion of the grain
mantles leads to observed enhancements of, e.g., \meth\ \citep{bachiller:l1157},
while the grain cores `sputter' refractive atoms such as Si and
Fe. Neutral-neutral reactions in the hot gas then transform these atoms into,
e.g., SiO, which is widely used as tracer of outflows \citep{jesus:sio}, and the
recently detected SiN and FeO molecules \citep{walmsley:feo,schilke:sin}.

\section{Galactic Variations in Cosmic-Ray Flux}
\label{sec:zeta}

The ionization fraction of molecular clouds determines the efficiency of
magnetic support against their gravitational collapse, and also sets the time
scale for ion-molecule chemistry. In star-forming regions, the bulk of the
matter is shielded against ultraviolet radiation, and cosmic rays are the main
ionization source. Only very close to embedded stars, photo-ionization plays a
role, as recent detections of CO$^+$ and SO$^+$ testify \citep{staeuber:ions}.
Cosmic rays influence molecular abundances not only through their total flux,
but also through their energy spectrum, in particular the ratio of H- to
He-ionizing particles \citep{wakelam:ions}.

Observations of molecular ions show significant variations in the cosmic-ray
ionization rate $\zeta$ within our Galaxy.  Submillimeter emission data of
\hcop\ toward a sample of seven high-mass star-forming regions at distances of
1--4\,kpc indicate $\zeta \sim 3\times 10^{-17}$\,\rs\ \citep{vdtak:zeta}. This
number is in good agreement with measurements of low-energy cosmic ray fluxes by
the Voyager and Pioneer spacecraft \citep{webber:zeta}. However, observations of
\dcop\ in nearby (0.1\,kpc) starless molecular cores indicate an ionization rate
reduced by a factor of $\sim$10 from this value \citep{caselli:zeta}.
On the other hand, 10$\times$ larger ionization
rates are found from \hhhp\ absorption data on the nearby (0.3\,kpc) $\zeta$~Per
cloud (\citealt{mccall:zper}; \citealt{lepetit:h3+}), and especially toward the
Sgr~A region near the Galactic center \citep{oka:sgra}. Enhanced $\zeta$-values
near the Galactic center are also reported from \hhhop\ observations of the
Sgr~B2 cloud \cite{vdtak:sgrb2}, but the derived ionization rate is lower than
that from \hhhp.

At least two effects appear responsible for the observed variations. First, the
cosmic-ray flux appears to decrease by a factor of $\sim$10 from the inner to
the outer Galaxy, as corroborated by synchrotron, X-ray and $\gamma$-ray data
\citep{yusef-zadeh:gc}. Second, scattering of cosmic rays off plasma waves
appears to cause the difference between diffuse and dense clouds. This process
is more efficient in denser clouds with stronger magnetic fields, in agreement
with the observations. However, other mechanisms may also play a
role. Observational estimates of $\zeta$ in regions with known magnetic field
strengths will help to make progress on this front.
%

The recent detections of interstellar and circumstellar C$_4$H$^-$, C$_6$H$^-$
and C$_8$H$^-$ mark the discovery of negative ions in space
(\citealt{mccarthy:c6h-}; \citealt{remijan:c8h-}; \citealt{cernicharo:c4h-};
\citealt{bruenken:c8h-}; \citealt{sakai:c6h-}). 
The large electron affinities of hydrocarbon chains makes the anionic species
almost as abundant as the neutral species \citep{herbst:anions,millar:ions}.
The total abundances only imply a small shift of negative charge, so that the
above estimates of the ionization rates of star-forming regions are not affected.
The negative ions are useful though, because combined with measurements of the
\hi\ 21\,cm line, the abundance ratios C$_n$H$^-$/C$_n$H may be used to estimate
the electron abundances in dark clouds \citep{flower:ions}. 

\section{Effects of Protostellar Irradiation}
\label{sec:x-ray}

During their main sequence phase, high-mass stars emit $\sim$10$^{-7}$ of their
luminosity in the form of X-rays, which originate in wind shocks. X-ray
observations of star-forming regions mainly probe the low-mass population, which
emits X-rays due to magnetic and accretion activity (see review by
\citealt{feigelson:ppv}). The onset of X-ray emission from high-mass stars is
hidden from our view, because of obscuration by the surrounding
material. However, the protostellar X-ray emission may be probed indirectly
through its effect on the chemistry of its molecular envelope.

\citet{benz:2591} have imaged the CS and SO submillimeter line emission from the
young high-mass star AFGL 2591 with the SubMillimeter Array. The data show a
pronounced `jump' in the SO abundance by a factor of $\sim$100 at a radius of
$\sim$1000\,AU. Model calculations by \citet{staeuber:xray} show that such a
jump is evidence for protostellar X-ray emission. 
Models with ice evaporation but without X-rays do not fit the data.
The derived $L_X$ is
$\sim$10$^{-6}$ of the total luminosity of AFGL 2591, which is somewhat higher
than for main sequence objects. Possibly the stellar winds of high-mass
protostars are stronger than those of main sequence stars, or additional X-ray
emission is generated in the interaction of the wind with the surrounding
envelope. 

\section{Prospects}
\label{sec:future}

The year 2008 will see the launch of ESA's \textit{Herschel} satellite, and
first data are expected in early 2009. Unhindered by the Earth's atmosphere,
this mission will make a major and unique contribution to astrochemistry,
especially with its spectrometer HIFI which covers the 480--1250 and
1410--1910\,GHz ranges at a resolution better than 1\,\kms. The highlights of
HIFI science will be, from an astrochemical point of view:

\begin{itemize}
\item unbiased spectral surveys of several Galactic star-forming
  regions, which provide inventories of their molecular composition;
\item large-scale maps of the \hho\ emission from dense clouds, and
  detailed multi-line studies of the \hho\ abundance distribution in
  star-forming regions;
\item precise measurements of the \oo\ abundance in dense clouds,
  PDRs and other environments;
\item measure the abundances of interstellar hydrides such as NH, a cornerstone
  of nitrogen chemistry (which is poorly known because N$^0$ does not have fine
  structure lines and \nn\ has no rotational lines.);
\item make an inventory of the major carbon and oxygen species in external
  galaxies, to study chemistry under more extreme conditions (including
  metallicity) than our Galaxy offers.
\end{itemize}

And just when the Herschel data will have been digested, ALMA operations will
get in full swing.  One byproduct will be lots of `accidental' astrochemists,
who find their submillimeter spectra full of unexpected spectral lines around
the line they were interested in. This reviewer hopes that these researchers
will evolve one day into `active' astrochemists.

\acknowledgements 
The author thanks Malcolm Walmsley, Ted Bergin and Chris McKee for useful
comments and suggestions.

\bibliographystyle{aa}
\bibliography{vdtak}



\end{document}